\documentclass[%
preprint,
 showpacs,preprintnumbers,
 nofootinbib,
 amsmath,amssymb,
 aps,
pdftex
]{revtex4-1}

\usepackage{graphicx}

\usepackage{bm}
\usepackage{amsmath}
\usepackage{ulem}
\usepackage[usenames]{color}
\usepackage{comment}

\begin{document}



\newcommand{\bbra}[1]{\langle\!\langle {#1} |}     
\newcommand{\kket}[1]{| {#1} \rangle\!\rangle}     
\newcommand{\rbra}[1]{( {#1} |}     
\newcommand{\rket}[1]{| {#1} )}     
\newcommand{\rdket}[1]{|\!| {#1} )}     
\newcommand{\rrket}[1]{| {#1} ))}     
\newcommand{\dbra}[1]{\langle {#1} |\!|}     
\newcommand{\rdbra}[1]{( {#1} |\!|}     
\newcommand{\dket}[1]{|\!| {#1} \rangle}     
\newcommand{\maru}[1]{\breve{#1}} 
\newcommand{\wtilde}[1]{\widetilde{#1}} 
\newcommand{\lsim}{{\stackrel{<}{\sim}}}
\newcommand{\gsim}{{\stackrel{\displaystyle >}{\raisebox{-1ex}{$\sim$}}}}
\newcommand{\lal}{\langle\!\langle}
\newcommand{\rar}{\rangle\!\rangle}
\newcommand{\wb}[1]{\overline{#1}}
\newcommand{\vect}[1]{\overrightarrow{#1}}
\newcommand{\ovl}[1]{\overline{#1}}
\newcommand{\braket}[1]{\langle{#1}\rangle}
\def\beq{\begin{eqnarray}}
\def\eeq{\end{eqnarray}}
\def\bsub{\begin{subequations}}
\def\esub{\end{subequations}}
\def\beq{\begin{eqnarray}}
\def\eeq{\end{eqnarray}}
\def\bsub{\begin{subequations}}
\def\esub{\end{subequations}}
\def\b{\begin{equation}}
\def\bs{\begin{split}}
\def\es{\end{split}}
\def\e{\end{equation}}

\title{
Hybrid stars from a three-flavor NJL model with two kinds of tensor condensates
}
\author{Masatoshi Morimoto}%
\email{b18d6a05@s.kochi-u.ac.jp}
\affiliation{Graduate School of Integrated Arts and Science, Kochi University, Kochi 780-8520, Japan}

\author{Yasuhiko Tsue}
\email{tsue@kochi-u.ac.jp}
\affiliation{Department of Mathematics and Physics, Kochi University, Kochi 780-8520, Japan}

\author{Jo\~{a}o da Provid\^{e}ncia}%
\author{Constan\c{c}a Provid\^{e}ncia}
\affiliation{CFisUC, Departamento de F\'{i}sica, Universidade de Coimbra, 3004-516 Coimbra, Portugal}

\author{Masatoshi Yamamura}
\affiliation{Department of Pure and Applied Physics, Faculty of Engineering Science, Kansai University,
Suita 564-8680, Japan}




\begin{abstract}
To obtain the equation of state of quark matter and construct hybrid stars,
we calculate the thermodynamic potential in the three-flavor Nambu-Jona-Lasinio model
including the tensor-type four-point interaction and the
Kobayashi-Maskawa-'t Hooft interaction.
To construct the hybrid stars, it is necessary to
impose the $\beta$ equilibrium and charge neutrality conditions on the system.
It is shown that tensor condensed phases appear at large chemical potential.
Under the possibility of the existence of the tensor condensates,
the relationship between the radius and mass of hybrid stars is estimated.
\end{abstract}



\maketitle

\section{Introduction}

One of recent interests is to reveal the phase structure of
the world governed by quantum chromodynamics (QCD).
In the region of high temperature and zero quark chemical potential,
the numerical simulation by using the lattice QCD gives a useful
information about the phase structure.
However, in the region of low temperature and large quark chemical potential,
the lattice simulation is still not working.
In that region, it has been remarked that various phases \cite{FH}
may appear such as the color superconducting phase
\cite{ARW,IB,CFL}, the quarkyonic phase,\cite{McL}
the inhomogeneous chiral condensed phase,\cite{NT}
the quark ferromagnetic phase,\cite{Tatsumi} the color-ferromagnetic phase,\cite{Iwazaki}
the spin polarized phase due to the axial vector interaction\cite{NMT,TMN,Maedan,Morimoto1,Morimoto2}
or due to the tensor interaction
\cite{BJ,IJMP,oursPTP,oursPTEP1,oursPTEP2,oursPTEP3,oursPTEP4,oursPR,Ferrer,MT,MNYY,India,Kagawa} and so on.
The situation with large chemical potential and low temperature may be realized
in the inner core of compact stars such as neutron stars and magnetars\cite{magnetar1,magnetar2}.
Especially, magnetars show a very strong magnetic field.
The spin polarization in quark matter may be the origin of the strong  magnetic field of magneers.
\cite{HL}

For estimating properties of high density quark matter,
the Nambu-Jona-Lasinio (NJL) model \cite{NJL}
is widely used \cite{Klevansky,HK,Buballa}
as one of the effective models of QCD.
In a previous paper\cite{India} and in our recent paper\cite{Kagawa}, the spin polarization
due to the tensor condensate has been investigated in the case of the
three-flavor NJL model with the Kobayashi-Maskawa-'t Hooft interaction.\cite{KM,determinant}
Also, properties of the compact star is investigated in the case of the
two-flavor NJL model with tensor interaction\cite{oursPR2}.
Thus, the investigation of the compact stars with the three-flavor quark matter in the inner core
of the compact stars such as hybrid stars is left.
So, in this paper, we investigate the properties of quark matter with the beta equilibrium and charge neutrality
conditions, which are necessary to construct the compact stars, by
using the three-flavor NJL model with the Kobayashi-Maskawa-'t Hooft interaction.

In Sec.\ref{sec:NJLmodel}, we introduce the three-flavor NJL model with the tensor
interaction and then calculate the effective potential.
In Sec.\ref{sec:numericalresults},
we discuss the behaviors of the tensor condensates obtained numerically under the beta equilibrium and the
charge neutrality conditions.
In Sec.\ref{sec:Hybridstar}, we construct hybrid stars and discuss numerical results.
The last section is devoted to a summary and concluding remarks.

\section{Recapitulation of mean field approximation for the Nambu-Jona-Lasinio model
with tensor-type four-point interaction between quarks \label{sec:NJLmodel}}

Let us start from the three-flavor
NJL model with tensor-type \cite{IJMP,oursPTP}
four-point interactions between quarks.
The Lagrangian density can be expressed as
\begin{align}
	\label{2-1}
\mathcal L&= \mathcal{L}_0+\mathcal{L}_m+\mathcal{L}_S
+\mathcal{L}_{T}+\mathcal{L}_{\rho}+\mathcal{L}_{e}+\mathcal{L}_{\rho_e}+\mathcal{L}_{D}
\\
	&\mathcal{L}_0={\bar \psi}i\gamma^\mu \partial_{\mu}\psi
		\nonumber \\
	&\mathcal{L}_m=-{\bar \psi}\vec m_0\psi
		\nonumber \\
	&\mathcal{L}_S=\frac{G_s}{2}\sum^8_{a=0}[({\bar \psi}\lambda_a\psi)^2+({\bar \psi}i\lambda_a\gamma_5\psi)^2]
		\nonumber \\
	&\mathcal{L}_{T}=-\frac{1}{2}\frac{G_T}{4}\sum^8_{a=0}[({\bar \psi}\gamma^\mu\gamma^\nu\lambda_a\psi)^2+({\bar \psi}i\gamma_5\gamma^\mu\gamma^\nu\lambda_a\psi)^2]
		\nonumber\\
	&\mathcal{L}_{\rho}=\mu_q{\psi^\dagger}\psi
		\nonumber \\
	&\mathcal{L}_{e}={\bar\psi_e}i\gamma^\mu\partial_\mu\psi_e
		\nonumber
\\
	&\mathcal{L}_{\rho_e}=\mu_e \left(\psi^\dagger_e \psi_e  - \frac{2}{3}\psi^\dagger_u \psi_u + \frac{1}{3}\psi^\dagger_d \psi_d  + \frac{1}{3}\psi^\dagger_s \psi_s \right)
		\nonumber
\\
	&\mathcal{L}_{D}=G_D\left[\text{det}\bar\psi(1-\gamma_5)\psi + \text{det}\bar\psi(1+\gamma_5)\psi\right]
\ .
\nonumber
\end{align}
Here,
$\vec m_0$ represents a current quark mass matrix in flavor space
as follows :
\begin{align}\label{2-2}
	\vec m_0&= \text{diag}\left( m_u,m_d,m_s \right) \ ,
\end{align}
where diag. means the diagonal matrix elements and the all off-diagonal elements are zero.
Here, ${\cal L}_T$, which is newly introduced in the original NJL model, represents a tensor-type
four-point interaction between quarks in the three-flavor case which preserves the chiral symmetry.
Also, ${\cal L}_D$ represents the so-called Kobayashi-Maskawa-'t Hooft or the determinant
interaction term
which leads to the six-point interaction between quarks in the three-flavor case.
This term resolves the problem of the $U_A(1)$-anomaly.
Further, $\mathcal{L}_{e}$ represents a kinetic term of electron.
In order to deal with the system of finite density,
$\mathcal{L}_{\rho}$ and $\mathcal{L}_{\rho_e}$  represent density terms for quarks and electron,
in which $\mu_q$ and $\mu_e$ correspond to the quarks and the electron chemical potentials,
respectively.
Each quark chemical potential is defined by
\begin{align}
	\mu_u &= \mu_q - \frac{2}{3} \mu_e \nonumber \\
	\mu_d &= \mu_q + \frac{1}{3} \mu_e = \mu_u + \mu_e \nonumber \\
	\mu_s &= \mu_d \ ,
\end{align}
which imply $\beta$ equilibrium condition.

Hereafter, we treat the above model within the mean field approximation,
in which we ignore non-diagonal components of the condensates in a flavor space.
Therefore, terms in the summation over $a$ 
are restricted to the diagonal entries
with $a=0, 3$ and 8
in ${\cal L}_S$ :
\begin{align}\label{2-3}
	\sum^{8}_{a=0}[({\bar \psi}\lambda_a\Gamma\psi)^2] &
\longrightarrow
 \sum_{a=0,3,8}[({\bar \psi}\lambda_a\Gamma\psi)^2]\nonumber \\
	&\quad
=\frac{2}{3}\left[\left(\bar u \Gamma u +\bar d \Gamma d +\bar s\Gamma s\right)\right]^2
	+\left[ \left(\bar u\Gamma u - \bar d\Gamma d\right)\right]^2 \nonumber\\
	& \qquad +\frac{1}{3}\left[\left(\bar u\Gamma u +\bar d\Gamma d -2\bar s\Gamma s\right)\right]^2 \nonumber \\
	&\quad
= 2(\bar u\Gamma u)^2 +2(\bar d\Gamma d)^2+2(\bar s\Gamma s)^2 \ .
\end{align}
Here, $\Gamma$ means products of any gamma matrices or unit matrix.
Also, in the determinant interaction term, ${\cal L}_D$,
the same approximation is adopted, namely, the off-diagonal matrix elements in
the flavor space are omitted:
\begin{align}\label{2-4}
	&\text{det}\bar\psi\left(1-\gamma_5\right)\psi + \text{det}\bar\psi\left(1+\gamma_5\right)\psi \nonumber \\
	&\longrightarrow
 \text{det}
	\begin{pmatrix}
		\bar u(1-\gamma_5)u & 0 & 0 \\
		 0 & \bar d(1-\gamma_5)d &0 \\
		0 & 0 &\bar s(1-\gamma_5)s
	\end{pmatrix}
\nonumber\\
&\qquad + \text{det}
	\begin{pmatrix}
		\bar u(1+\gamma_5)u & 0 & 0 \\
		0 & \bar d(1+\gamma_5)d &0 \\
		0 & 0 &\bar s(1+\gamma_5)s
	\end{pmatrix} \nonumber\\
	&= 2(\bar uu) (\bar dd) (\bar ss)
\nonumber\\
&\quad
+ 2(\bar uu) (\bar d\gamma_5d) (\bar s\gamma_5s)+2(\bar u\gamma_5u) (\bar dd) (\bar s\gamma_5s)+2(\bar u\gamma_5u) (\bar d\gamma_5d) (\bar ss) \ .
\end{align}
Secondly, in order to consider the spin polarization under the mean field approximation,
the tensor condensate $\braket{{\bar q}\gamma^1\gamma^2 q}$ and
$\braket{{\bar q}\gamma^2\gamma^1 q}$ are considered
in ${\cal L}_T$
because
$\gamma^1\gamma^2=i\Sigma_3$.
Here,
\beq\label{2-5}
\Sigma_3=-i\gamma^1\gamma^2=
\begin{pmatrix}
		\sigma_3 & 0  \\
		0 & \sigma_3
	\end{pmatrix} \ ,
\eeq
where $\sigma_3$ represents the third component of the Pauli matrix.
Thus, we consider two tensor condensates under the mean field approximation as
\beq\label{2-6}
& &F_3= -G_T\braket{{\bar \psi}\Sigma_3\lambda_3\psi}\ , \nonumber\\
& &F_8= -G_T\braket{{\bar \psi}\Sigma_3\lambda_8\psi}\ .
\eeq
For each quark flavor, the tensor condensates are reexpressed as
\beq\label{2-7}
& &F_u=F_3+\frac{1}{\sqrt{3}}F_8\ , \nonumber\\
& &F_d=-F_3+\frac{1}{\sqrt{3}}F_8\ , \nonumber\\
& &F_s=-\frac{2}{\sqrt{3}}F_8\ .
\eeq
Of course, the chiral condensates $\langle {\bar q}q\rangle$ should be taken into account.
We introduce the dynamical quark masses ${\cal M}_f$
without the determinant interaction term
by using the chiral condensates as
\beq\label{2-8}
& &{\cal M}_u=-2G_s\langle {\bar u}u\rangle\ , \nonumber\\
& &{\cal M}_d=-2G_s\langle {\bar d}d\rangle\ , \nonumber\\
& &{\cal M}_s=-2G_s\langle {\bar s}s\rangle\ .
\eeq
These expressions are only valid if the mixing term is not considered.

Thus, under the mean field approximation, the Lagrangian density (\ref{2-1}) reduces to
\begin{align}
\label{eq:L_{MF}}
\mathcal L_{MF}=&{\bar \psi}(i\gamma^\mu \partial_\mu -\vec M - \vec F \Sigma_3 + \vec\mu \gamma^0)\psi \nonumber \\
&-\sum_f\frac{\mathcal M_f^2}{4G_s} -\frac{F_3^2}{G_T} - \frac{F_8^2}{G_T} +\frac{1}{2}\frac{G_D}{G_s^3} \mathcal M_u \mathcal M_d \mathcal M_s\ \nonumber \\
&+\bar \psi_e\left(i\gamma^\mu\partial_\mu + \mu_e\gamma^0   \right)\psi_e\ ,
\end{align}
where $f=u,\ d$ or $s$ and
\begin{align}
&\Sigma_3=-i\gamma^1\gamma^2=
\left(
\begin{array}{cc}
\sigma_3 & 0 \\
0 & \sigma_3
\end{array}
\right)
	\nonumber\\
	&\vec M =\text{diag.}\left(m_u + \mathcal M_u -\frac{G_D}{2G_s^2}\mathcal M_d \mathcal M_s \ ,\ m_d + \mathcal M_d  -\frac{G_D}{2G_s^2}\mathcal M_s  \mathcal M_u\ , \right.
	\nonumber\\
	&\qquad\qquad\qquad \left. \ m_s + \mathcal M_s -\frac{G_D}{2G_s^2}\mathcal M_u \mathcal M_d  \right)
	\nonumber\\
	&\quad =\text{diag.}(M_u,\ M_d,\ M_s) \nonumber\\
	&\vec F=
\text{diag.}\left(F_u,\ F_d,\ F_s\right)
	\nonumber \\
	&\vec \mu = \text{diag.}\left(\mu_u\ ,\ \mu_d\ ,\ \mu_s\right)
	. \label{eq:L_{MF}2}
\end{align}
Here, ${\vec M}$ represents the constituent quark mass matrix with the flavor mixing due to the determinant interaction term.
For example, $M_u=m_u + {\cal M}_u-({G_D}/{2G_s^2}){\cal M}_d{\cal M}_s$ represents the constituent quark mass
for $u$-quark,
which consists of the current quark mass for $u$-quark, $m_u$, dynamical quark mass for $u$-quark, ${\cal M}_u$
originated from the dynamical chiral symmetry breaking or quark condensate in (\ref{2-8}) and
the effect of the flavor mixing due to the Kobayashi-Maskawa-'t Hooft term with $G_D$ being non-zero value.

The Hamiltonian density can be obtained from the mean field Lagrangian density as
\begin{align}
  \label{2-11}
	\mathcal{H}_{MF} = &\bar\psi \left(i \bm \gamma \cdot \bm\nabla + \vec M + \vec F \Sigma_3 \right) \psi
- \psi^\dagger \vec \mu  \psi
	\nonumber \\
	&+\sum_f\frac{\mathcal M_f^2}{4G_s} + \frac{F_3^2}{G_T} + \frac{F_8^2}{G_T} - \frac{1}{2}\frac{G_D}{G_s^3}\mathcal M_u\mathcal M_d\mathcal M_s
	\nonumber \\
	&+\bar \psi_e\left(i\bm \gamma \cdot \bm\nabla  \right)\psi_e - \mu_e\psi_e^\dagger\psi_e
	\nonumber \\
  =& \psi^\dagger {\hat h_q} \psi - \psi^\dagger \vec \mu \psi + \psi_e^\dagger \hat h_e \psi_e - \mu_e\psi_e^\dagger\psi_e
  \nonumber \\
  &+\sum_f\frac{\mathcal M_f^2}{4G_s} + \frac{F_3^2}{G_T} + \frac{F_8^2}{G_T} - \frac{1}{2}\frac{G_D}{G_s^3}\mathcal M_u\mathcal M_d\mathcal M_s \ .
\end{align}
Here, ${\hat h}_q$ and ${\hat h}_e$ mean the single-particle Hamiltonians for single quark and electron, respectively,
which are explicitly written in
\begin{align}
	\hat h_q
	&\equiv \gamma^0 \left( i \bm \gamma \cdot \bm\nabla + \vec M + \vec F \Sigma_3 \right) \ , \nonumber \\
  \hat h_e &\equiv \gamma^0 \left( i\bm \gamma \cdot \bm\nabla \right) \ .
\end{align}
Let us derive the effective potential or the thermodynamic potential at zero temperature.
In order to obtain
the energy eigenvalues of single quark and electron, it is necessary to diagonalize ${\hat h}_q$ and ${\hat h}_e$,
the eigenvalues of which can be obtained easily as
\begin{align}
  \label{2-14}
	E_f &= \sqrt{p_z^2 + \left( \sqrt{p_x^2+p_y^2+M_f^2} + \eta F_f  \right)}
	\qquad\left( f=u,d,s \right) \ ,  \nonumber \\
  E_e&= \left| \bm p \right|\ ,
\end{align}
where $\eta=\pm 1$.

Thus, we can easily evaluate the thermodynamic potential with the energy eigenvalues.
The thermodynamic potential $\Phi$ can be expressed as
\begin{align}
  \label{2-15}
	\Phi &= \braket{\mathcal{H}_{MF}} \nonumber \\
	 &= \Phi_u + \Phi_d + \Phi_s + \Phi_e + \Phi_{MF}\ ,
\end{align}
where each term represents
\begin{align}
  \label{eq:Phi_f}
	\Phi_f
	&= \sum_{\eta,\alpha} \int \frac{d^3 p}{\left(2\pi\right)^3} \left( E_{f,\eta}-\mu_f \right)\theta\left( \mu_f - E_{f,\eta}\right) - \sum_{\eta,\alpha} \int \frac{d^3 p}{\left(2\pi\right)^3} E_{f,\eta}\ ,
	\nonumber \\
	\Phi_e
	&= 2 \int^\infty_{-\infty} \frac{d^3 p}{\left(2\pi\right)^3} \left( E_{e}-\mu_e \right)\theta\left( \mu_e - E_{e}\right)
	\nonumber \\
	\nonumber \\
	&=\frac{1}{\pi^2}\int^{\mu_e}_0 d \left|\bm p \right| \left(\left|\bm p \right|^3 - \mu_e\left|\bm p \right|^2 \right)
	\nonumber \\
	&=-\frac{\mu_e^4}{12\pi^2} \ ,
	\nonumber \\
	\Phi_{MF}
	&=\frac{\mathcal M_u^2 + \mathcal M_d^2 + \mathcal M_s^2}{4G_s} + \frac{F_3^2}{G_T} + \frac{F_8^2}{G_T} - \frac{1}{2}\frac{G_D}{G_s^3}\mathcal M_u\mathcal M_d\mathcal M_s \ .
\end{align}
Here, $\alpha$ represents the color degree of freedom whose summation leads to numerical factor $N_c\ (=3)$.
Here, $\theta(x)$ represents the Heaviside step function.
The first and second terms of $\Phi_f$ represent the positive-energy contribution of quarks and the vacuum contribution, respectively.

To determine the chiral condensates, $\mathcal M_u$ ,  $\mathcal M_d$ and $\mathcal M_s$ , and the tensor condensates
$F_3$ and $F_8$, the gap
equations are
demanded as
\beq\label{2-16}
\frac{\partial \Phi}{\partial {\cal M}_u} =
\frac{\partial \Phi}{\partial {\cal M}_d} =
\frac{\partial \Phi}{\partial {\cal M}_s} =
\frac{\partial \Phi}{\partial {F_3}} =
\frac{\partial \Phi}{\partial {F_8}} =0 \ .
\eeq

Through the thermodynamical relations, quark
number density for each flavor $\rho_f$ can be estimated by differentiating
the thermodynamic potential with respect to the quark chemical potential :
\begin{align}
	\rho_f&=-\frac{\partial\Phi}{\partial\mu_f}=-\frac{\partial\Phi_f}{\partial\mu_q} \ .
\end{align}
Also, total quark number density $\rho_q$ is written as
\begin{align}
    \rho_q = \rho_u + \rho_d +\rho_s \ .
\end{align}
Similarly, the electron number density can be obtained as
\begin{align}\label{eq:rho_e}
	\rho_e &= -\frac{\partial\Phi}{\partial\mu_e}= \frac{\mu_e^3}{3\pi^2} \ .
\end{align}
Under the charge neutrality condition
\begin{align}
	\label{eq:chargeneutrality}
	\frac{2}{3}\rho_u - \frac{1}{3}\rho_d- \frac{1}{3}\rho_s - \rho_e = 0 \ ,
\end{align}
the thermodynamic potential should take the minimum value.
In (\ref{eq:chargeneutrality}), each
factor corresponds to electric charge for quark with each flavor and electron.

For estimating inner structure of compact star,
we must calculate the pressure and energy density.
By the use of the thermodynamical relations,
pressure $P$ and energy density $\epsilon$ are calculated as
\begin{align}\label{P}
  P &= - \left(\Phi - \Phi_{(\mu_q=0)}\right)\ , \nonumber \\
  \epsilon &= - P + \mu_q \rho_q \ .
\end{align}
Here, we have renormalized $P$ with the value at $\mu_q=0$,
which leads to $P=0$ at $\mu_q=0$.

\section{Numerical results\label{sec:numericalresults}}

In this and the next sections, we give numerical results.
In this section, the behavior of the tensor condensates etc. at finite chemical potential
is shown.
%
\begin{table}[t]
\caption{Parameter sets of 3-flavor NJL model with the tensor interaction.}
\label{table:modelparameters}
\begin{center}
\begin{tabular}{c||c|c|c|c|c|c}
\hline
 & $m_u,\ m_d$& $m_s$ & $G_s$ & $G_T$ & $G_D$ & $\Lambda$\\
 & [/GeV] & [/GeV] & [/GeV$^{-2}$] & [/GeV$^{-2}$] &  [/GeV$^{-5}$]
&[/GeV]
\\ \hline
Model GT0 & 0.0055 & 0.1375 &$ 3.666/\Lambda^2 (\approx 9.2)$& $0$ &9.288/$\Lambda^5$ & 0.6314 \\
Model GT2.0 & 0.0055 & 0.1375 &$ 3.666/\Lambda^2 $& $2G_s$ &9.288/$\Lambda^5$ & 0.6314 \\
Model GT2.2 & 0.0055 & 0.1375 &$ 3.666/\Lambda^2 $& $2.2G_s$ &9.288/$\Lambda^5$ & 0.6314 \\
Model GT2.4 & 0.0055 & 0.1375 &$ 3.666/\Lambda^2 $& $2.4G_s$ &9.288/$\Lambda^5$ & 0.6314 \\
Model GT2.6 & 0.0055 & 0.1375 &$ 3.666/\Lambda^2 $& $2.6G_s$ &9.288/$\Lambda^5$ & 0.6314 \\
\hline
\end{tabular}
\end{center}
\end{table}

The model parameter sets are summarized in Table \ref{table:modelparameters}.
In these parameters, the tensor interaction strength $G_T$ is taken
as a free parameter in our consideration,
while the value of $G_T$ may be estimated by the vacuum properties of pion and $\rho$ meson as
in Ref.\cite{Jaminon:2002}.
However,
this parameter could not be determined exactly by the experimental data of a certain physical quantity.
The tensor-type interaction may be derived
from a two-gluon exchange interaction\cite{oursPTEP4}.
However, the NJL model cannot be derived from the QCD Lagrangian directly.
Thus, we adopt $G_T$ as a free parameter in this model.
If the tensor interaction term is derived by the Fierz transformation
of the scalar interaction term,
$G_s({\bar \psi}\psi)^2$,
the relationship $G_T=2G_S$ is satisfied.

When the thermodynamic potential in (\ref{2-15}) is calculated,
a regularization scheme is necessary
because the vacuum contribution in the second line in (\ref{2-15}) gives the divergent contribution.
Here, we adopt the three-momentum cutoff scheme and introduce the three-momentum cutoff $\Lambda$.

The values of the parameters used here except for $G_T$ are adopted following Ref. \cite{HK}, in which
the parameters, namely the three-momentum cutoff $\Lambda$,
the coupling constant $G_s$, $G_D$, and
the current quark (light quark and strange quark) masses,
are given so as to reproduce the pion decay constant,  the chiral condensate, pion and kaon masses and
eta and eta-prime meson masses.
The Kobayashi-Maskawa-'t Hooft term leads to the axial $U_A(1)$-anomaly.
Thus, by introducing the non-zero coupling constant $G_D$ for
the Kobayashi-Maskawa-'t Hooft term, the eta and eta-prime meson-mass difference
must also be reproduced.
The way of decision of the parameters is developed in Ref.\cite{HK} in three-flavor case in detail.

\subsection{Behavior of consistent quark masses}


\begin{figure}[t]
	\begin{center}
		\includegraphics[height=5cm]{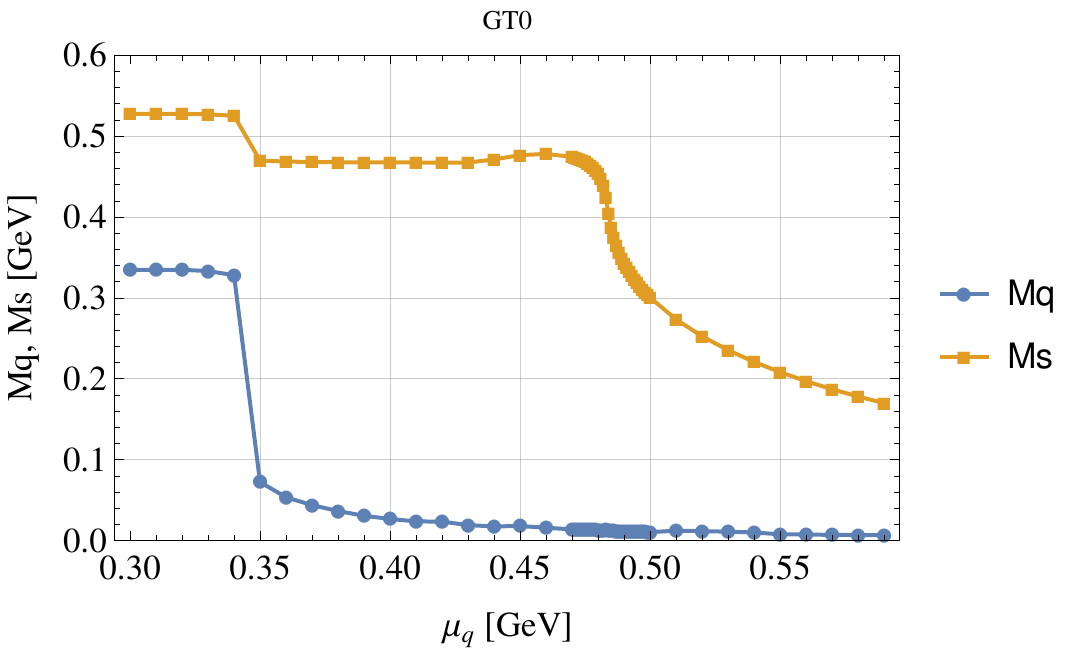}
	\end{center}
	\caption{Dynamical quark
masses $M_q$ and $M_s$ are depicted as a function of the quark chemical potential $\mu_q$ for model GT0.
The circle (lower curve) and square (upper curve) represent $M_q$ and $M_s$, respectively.
	\label{fig:MqMs0_mu}
}
\end{figure}

\begin{figure}[b]
	\begin{minipage}[t]{0.45\hsize}
	\begin{center}
		\includegraphics[height=4.0cm]{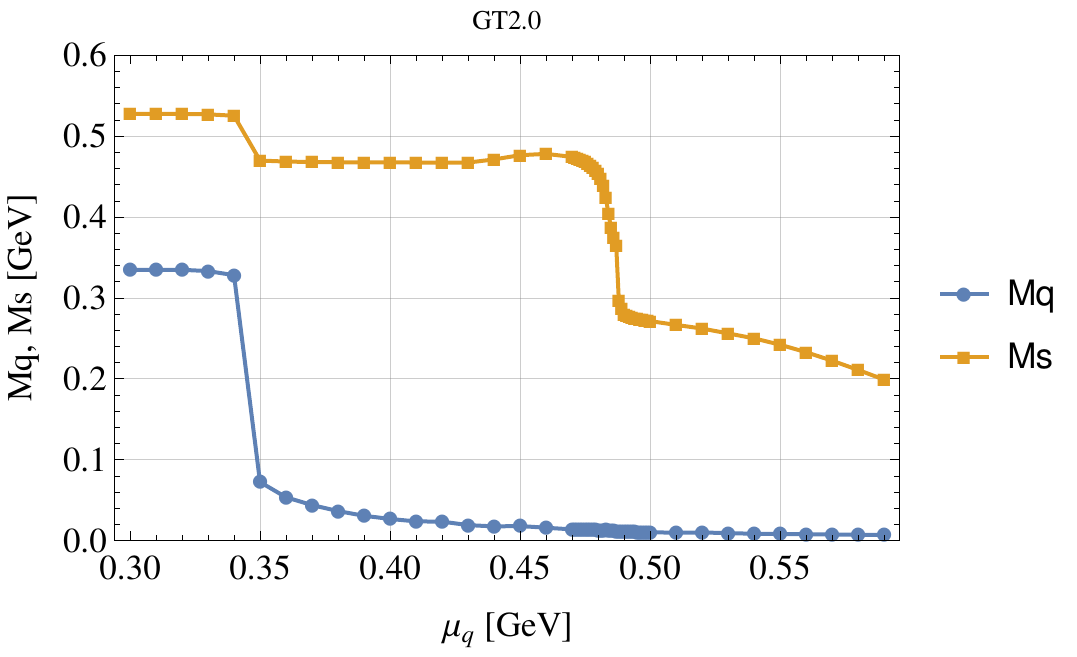}

	\end{center}
	\end{minipage}
\qquad
	\begin{minipage}[t]{0.45\hsize}
\begin{center}
		\includegraphics[height=4.0cm]{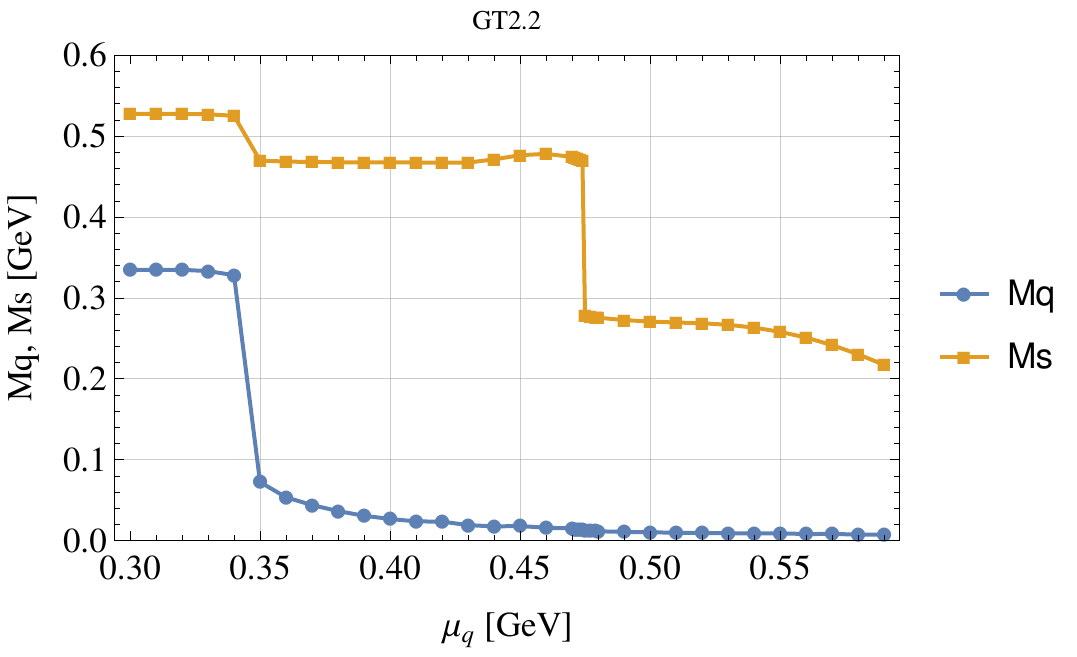}
	\end{center}
	\end{minipage}
	\\
	\begin{minipage}[t]{0.45\hsize}
	\begin{center}
		\includegraphics[height=4.0cm]{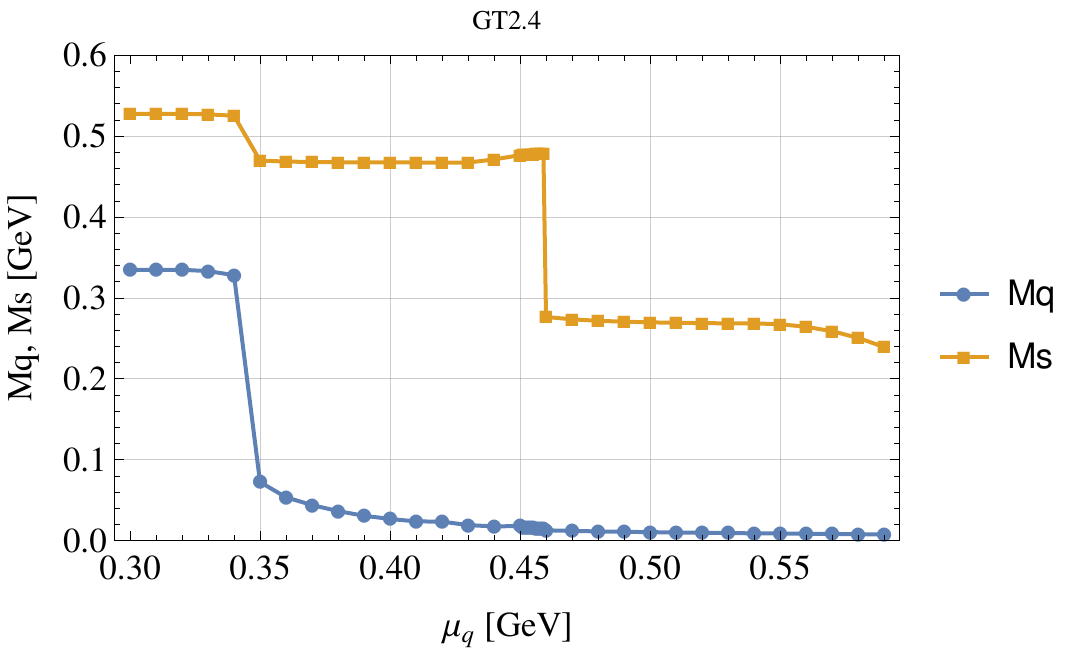}
	\end{center}
	\end{minipage}
\qquad
	\begin{minipage}[t]{0.45\hsize}
\begin{center}
		\includegraphics[height=4.0cm]{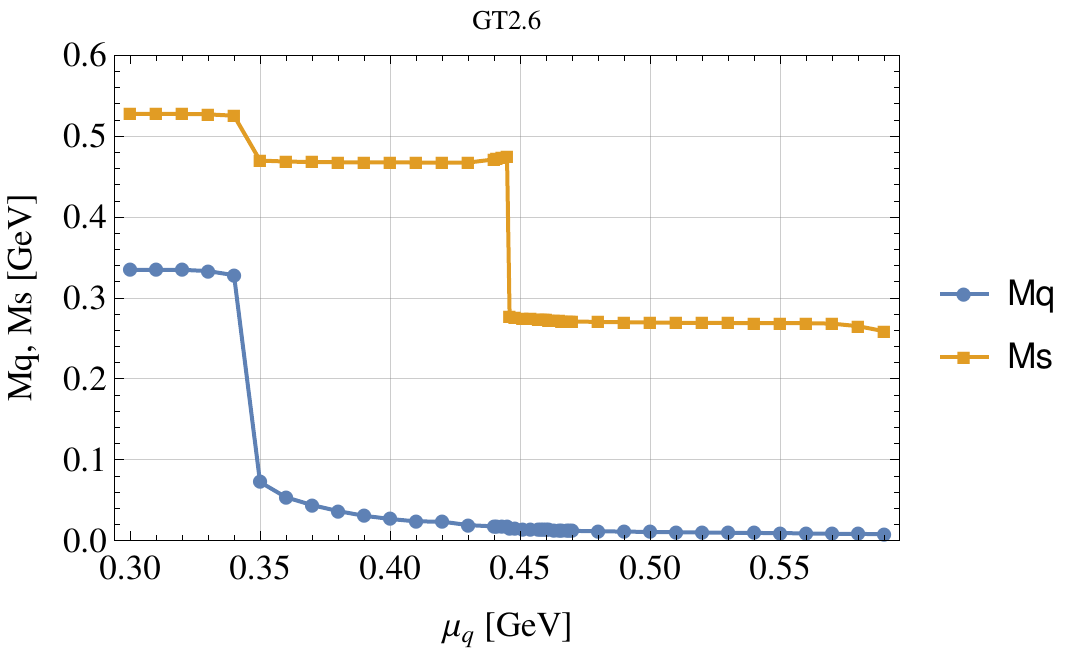}
	\end{center}
	\end{minipage}
	\caption{Dynamical quark masses $M_q$ and $M_s$ are depicted as a function of the quark chemical potential $\mu_q$ for each model of $G_T \neq 0$.
The circle (lower curves) and square (upper curves) represent $M_q$ and $M_s$, respectively.
	\label{fig:MqMs_mu}
}
\end{figure}

Figure \ref{fig:MqMs0_mu} shows the behavior of the consistent quark masses $M_q$ and $M_s$ for model GT0
without the tensor interaction.
In the region of $\mu_q \leq 0.34$ GeV, dynamical quark masses have values, namely
$M_q = 0.335$ GeV and $M_s =0.527$ GeV, by the dynamical chiral symmetry breaking.
At $\mu_q\approx 0.34$ GeV, the light quark masses decrease suddenly.
Then, the strange quark mass reveals a small gap,
which is originated from the flavor mixing.
Namely, the behaviors of the light quark masses have an effect on the behavior of the strange quark mass
by the flavor mixing.
At $\mu_q\approx 0.44$ GeV, the strange quark mass increases slightly.
This behavior is originated from the charge neutrality and $\beta$ equilibrium conditions.
Electron chemical potential takes the maximum value in this region.
The electron chemical potential is depicted in section \ref{subsec:rhoandmue}.
At $\mu_q \ge 0.47$ GeV, the strange quark mass decreases continuously.

Figure \ref{fig:MqMs_mu} shows the behavior of the constituent quark masses $M_q$ and $M_s$
for the models of GT2.0, GT2.2, GT2.4 and GT2.6 with finite value of the tensor interaction strength.
For the quark masses at the low chemical potential region,
these models reveal the same behavior as model GT0.
In the case of model GT2.0/GT2.2/GT2.4, at $\mu\approx 0.488/0.475/0.460 $ GeV $\equiv \mu_{cr_1}$,
the constituent quark mass of the strange quark, $M_s$, exhibits a mass gap, while $M_s$ exhibits only a small gap
in model GT2.0.
Also, model GT2.6 shows similar mass gap at $\mu\approx 0.446 $ GeV $\equiv \mu_{cr_{1'}}$.
The reason why these behavior of $M_s$ is shown is that
the tensor condensates appear at $\mu = \mu_{cr_1}$ and $\mu = \mu_{cr_{1'}}$.
Here, in the models GT2.0, GT2.2 and GT2.4, the tensor condensate $F_8$ only appears at $\mu=\mu_{cr_1}$.
On the other hand, in the model GT2.6, both the tensor condensates $F_8$ and $F_3$ appear at $\mu=\mu_{cr_{1'}}$
simultaneously.
Namely, the tensor condensates show different behaviors for model GT2.6 and others.
As for the tensor condensates, details are described in section \ref{subsec:tensor}.

\begin{figure}[b]
      \begin{tabular}{cc}
	\begin{minipage}[t]{0.45\hsize}
	\begin{center}
		\includegraphics[height=4.0cm]{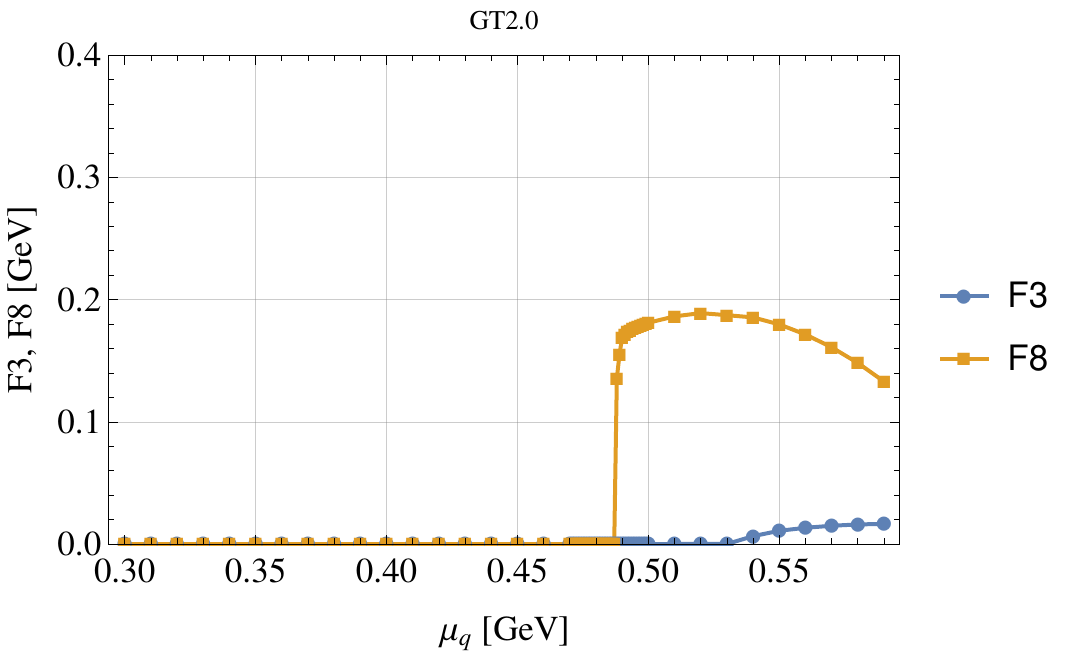}
	\end{center}
	\end{minipage}
\qquad
	\begin{minipage}[t]{0.45\hsize}
\begin{center}
		\includegraphics[height=4.0cm]{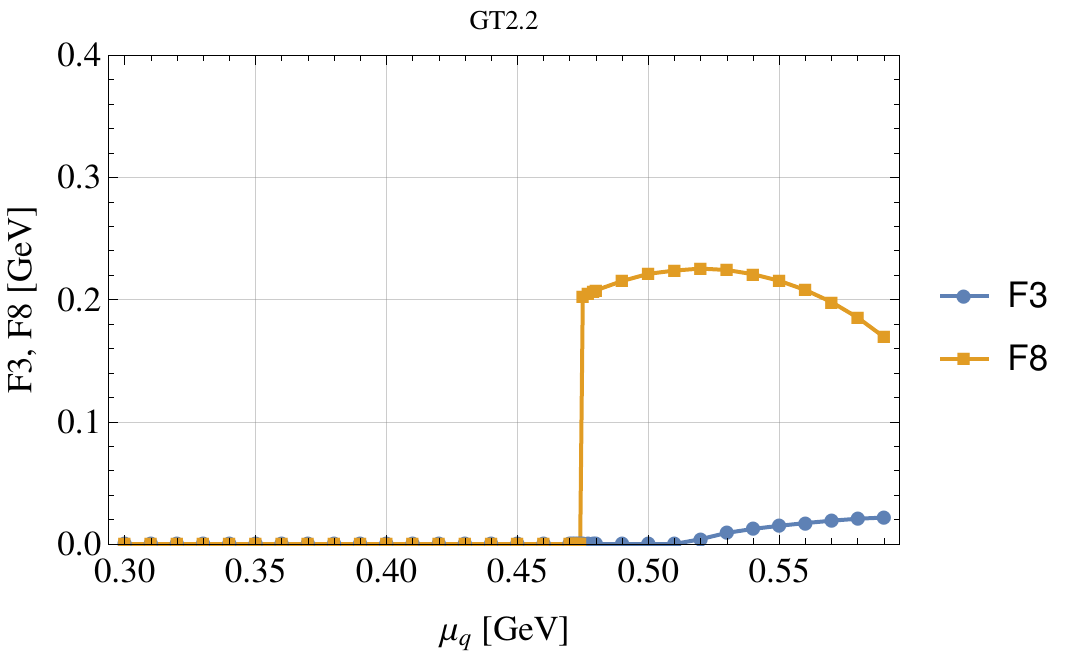}
	\end{center}
	\end{minipage}
	\\
	\begin{minipage}[t]{0.45\hsize}
	\begin{center}
		\includegraphics[height=4.0cm]{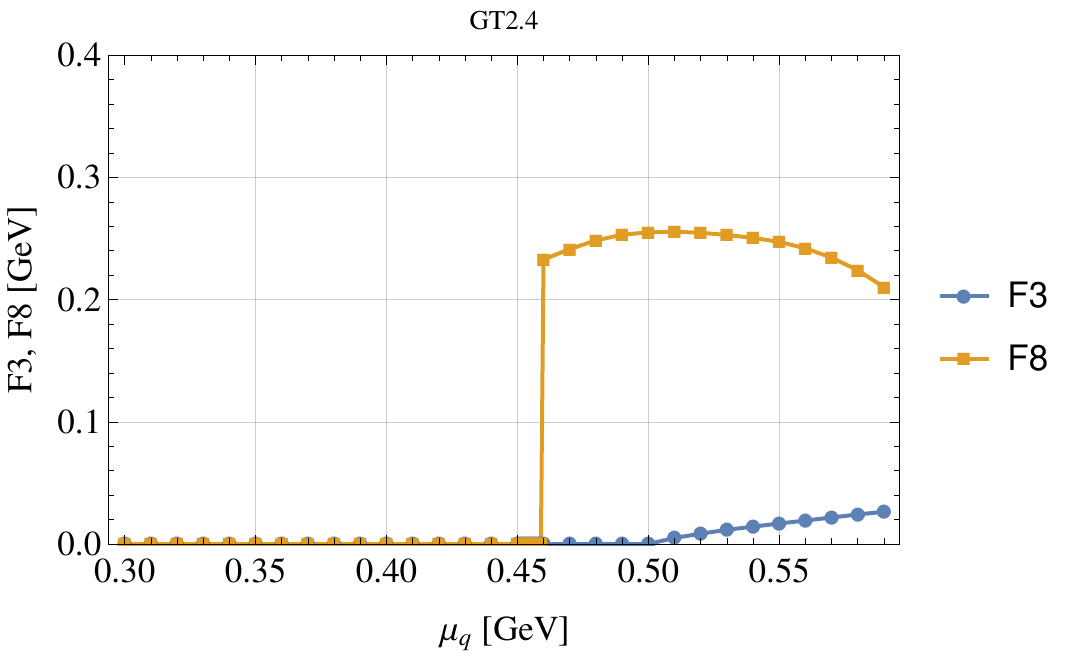}
	\end{center}
	\end{minipage}
\qquad
	\begin{minipage}[t]{0.45\hsize}
\begin{center}
		\includegraphics[height=4.0cm]{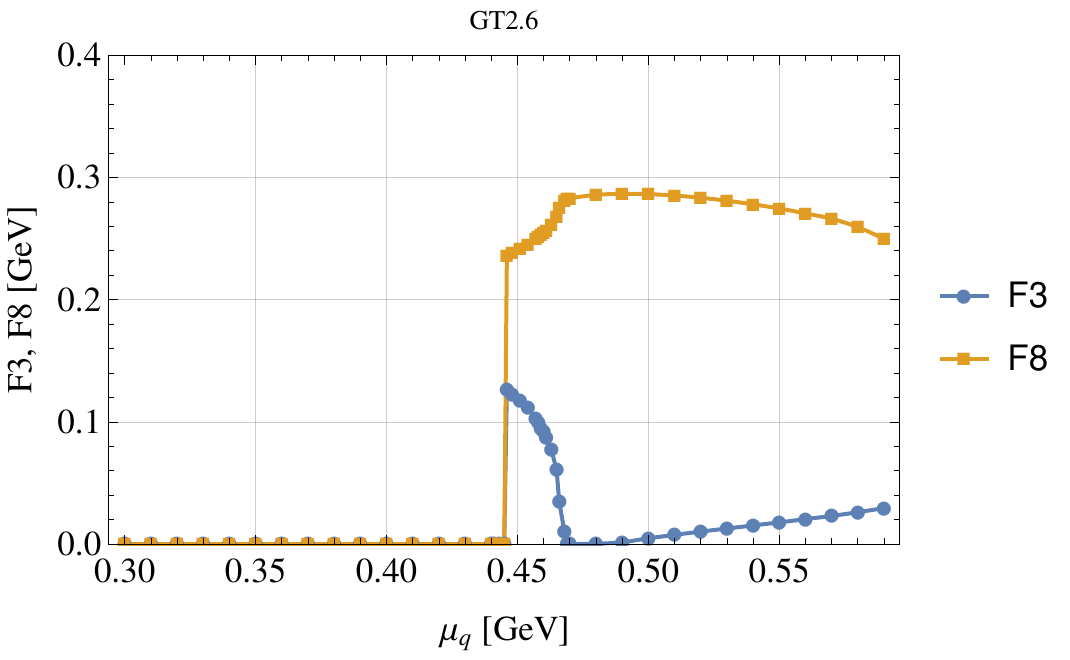}
	\end{center}
	\end{minipage}
  \end{tabular}
	\caption{Tensor condensates $F_3$ and $F_8$ are depicted as a function of the quark chemical potential $\mu_q$ for each model.
The circle (lower curves) and square (upper curves) represent $F_3$ and $F_8$, respectively.
	\label{fig:F3F8_mu}
}
\end{figure}

\subsection{Behavior of tensor condensates \label{subsec:tensor}}

Figure \ref{fig:F3F8_mu} shows the behavior of the two possible
tensor condensates $F_3$ and $F_8$ as a function of the quark chemical potential $\mu_q$.
In the model GT2.0/GT2.2/GT2.4, at $\mu_q
=\mu_{cr_1}$,
$F_8$ appears suddenly.
Also, $\mu_q \approx 0.54/0.52/0.51\equiv \mu_{cr_2}$ GeV, $F_3$ appears gradually.
Thus, two tensor condensates $F_3$ and $F_8$ coexist in the region of $\mu_q \geq \mu_{cr_2}$.
On the other hand, in the model GT2.6, both the tensor condensates, $F_3$ and $F_8$, appear simultaneously
at $\mu_q=\mu_{cr_{1'}}$.
In the region with $\mu_{cr_{1'}} \leq\mu_q\leq 0.469$ GeV $(\equiv \mu_{cr_{2'}}$), $F_3$ decreases.
Once $F_3$ disappears at $\mu_q = \mu_{{cr_2'}}$, but starts to increase again.
In the region $\mu_q \geq \mu_{cr_{2'}}$, each condensate shows the same behavior
as the other models of the region of $\mu_q \geq \mu_{cr_2}$.

\begin{figure}[b]
      \begin{tabular}{cc}
	\begin{minipage}[t]{0.45\hsize}
	\begin{center}
		\includegraphics[height=4.0cm]{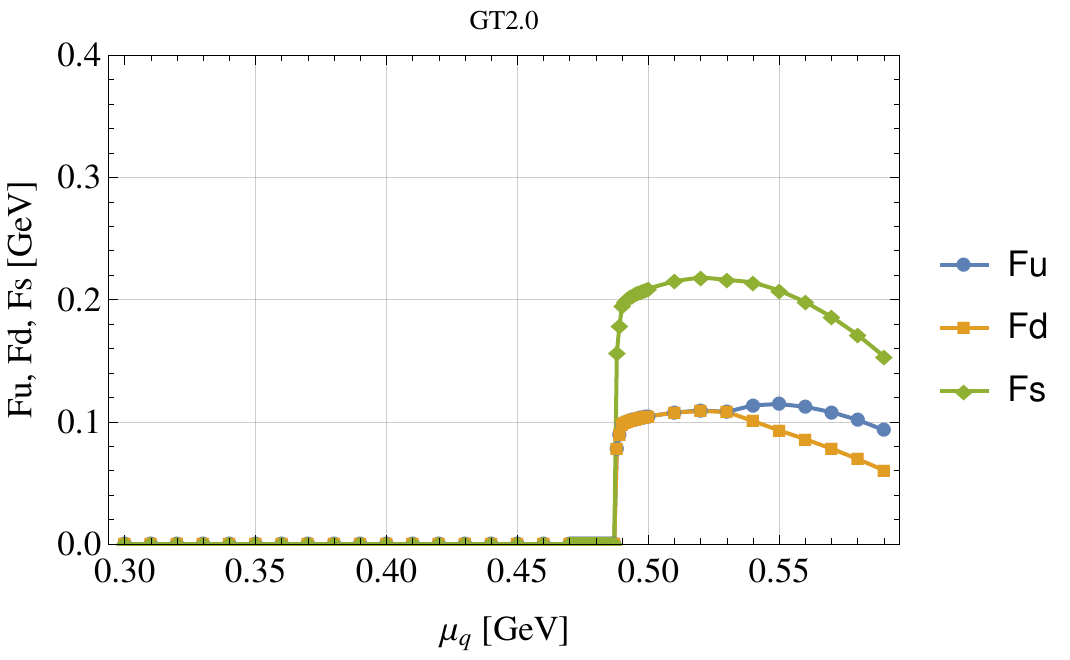}

	\end{center}
	\end{minipage}
\qquad
	\begin{minipage}[t]{0.45\hsize}
\begin{center}
		\includegraphics[height=4.0cm]{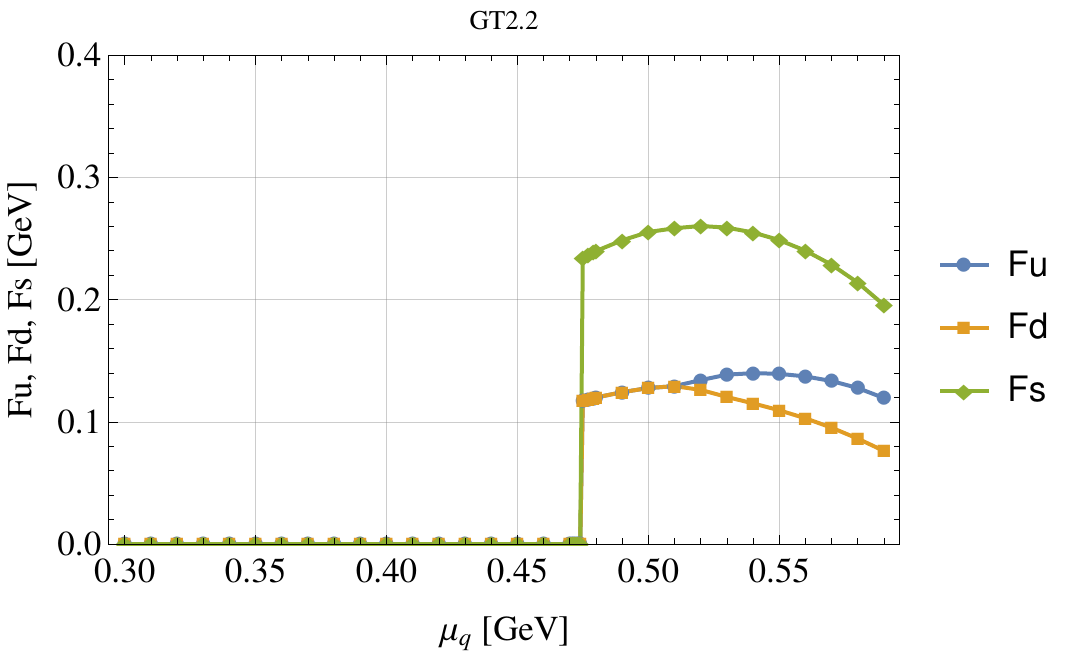}
	\end{center}
	\end{minipage}
	\\
	\begin{minipage}[t]{0.45\hsize}
	\begin{center}
		\includegraphics[height=4.0cm]{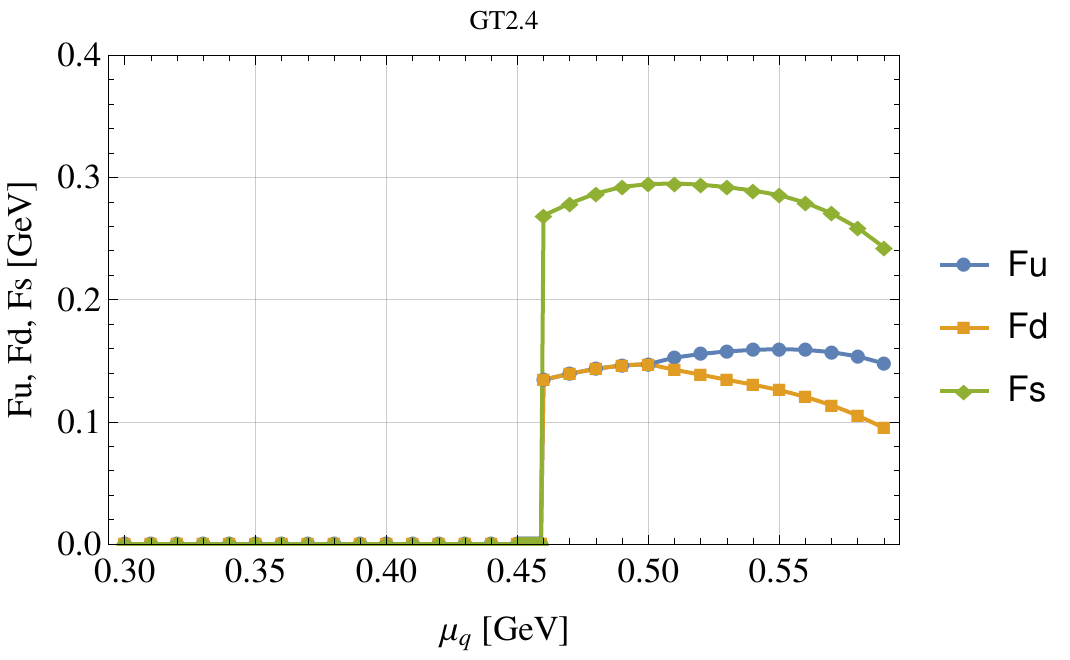}
	\end{center}
	\end{minipage}
\qquad
	\begin{minipage}[t]{0.45\hsize}
\begin{center}
		\includegraphics[height=4.0cm]{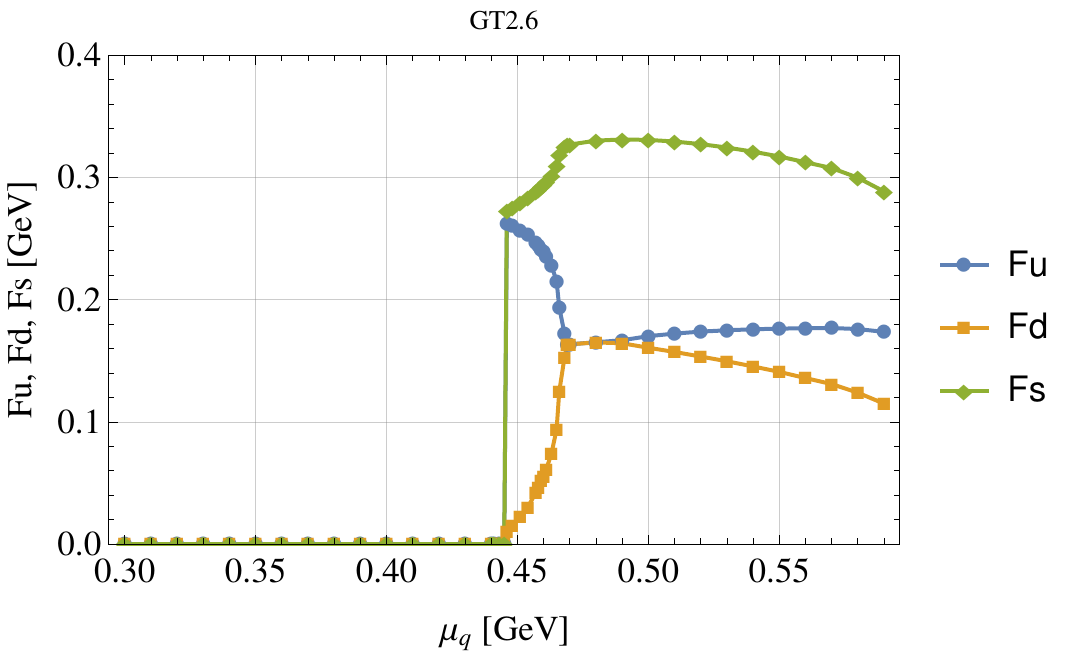}
	\end{center}
	\end{minipage}
  \end{tabular}
	\caption{Tensor condensates $F_u$, $F_d$ and $F_s$ are depicted as a function of the quark chemical potential $\mu_q$ for each model.
The circle (blue curves), square (orange curves) and diamond (green curves) represent $F_u$, $F_d$ and $F_s$, respectively.
	\label{fig:Fuds_mu}
}
\end{figure}

Figure \ref{fig:Fuds_mu} shows the behavior of the tensor condensates for each flavor, namely, $F_u$, $F_d$ and $F_s$,
in Eq. (\ref{2-7}).
In these figures, as for $F_s$, the absolute value $|F_s|$ is depicted because $F_s$ has a negative sign.
In the case of model GT2.0/GT2.2/GT2.4,
in the region of $\mu_{cr_1}\leq\mu_q\leq\mu_{cr_2}$,
only the tensor condensate $F_8$ appears.
Therefore, $F_u$ and $F_d$ have the same values.
After $F_3$ appears by increasing the quark chemical potential,
$F_u$ and $F_d$ begin to have different values.
In the case of model GT2.6,
at $\mu_q \approx \mu_{cr_{1'}}$,
the tensor condensates $F_u$ has a finite value, $F_u \neq 0$, but $F_d=0$.
The value of $F_u$/$F_d$ decreases/increases as $F_3$ decreases.
At $\mu \approx \mu_{cr_{2'}}$, $F_u$ and $F_d$ have the same values.
In the region of $\mu_q \geq \mu_{cr_{2'}}$, both the tensor condensates show the same behavior as the other models.

\subsection{Behaviors of quark number density $\rho$ and electron chemical potential $\mu_e$
\label{subsec:rhoandmue}}

\begin{figure}[b]
	\begin{center}
		\includegraphics[height=5.0cm]{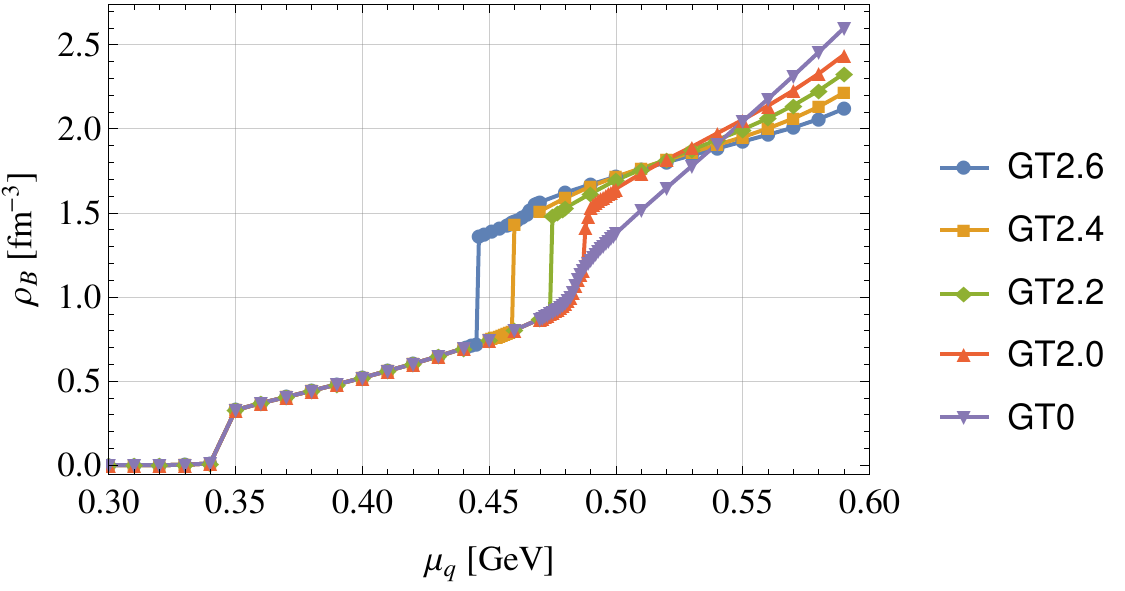}
	\end{center}
	\caption{Baryon number density is depicted as a function of the quark chemical potential $\mu_q$ for each model.
			\label{fig:rhoB-mu}
}
\end{figure}

\begin{figure}[t]
	\begin{center}
	\begin{minipage}[t]{0.45\hsize}
		\includegraphics[height=3.5cm]{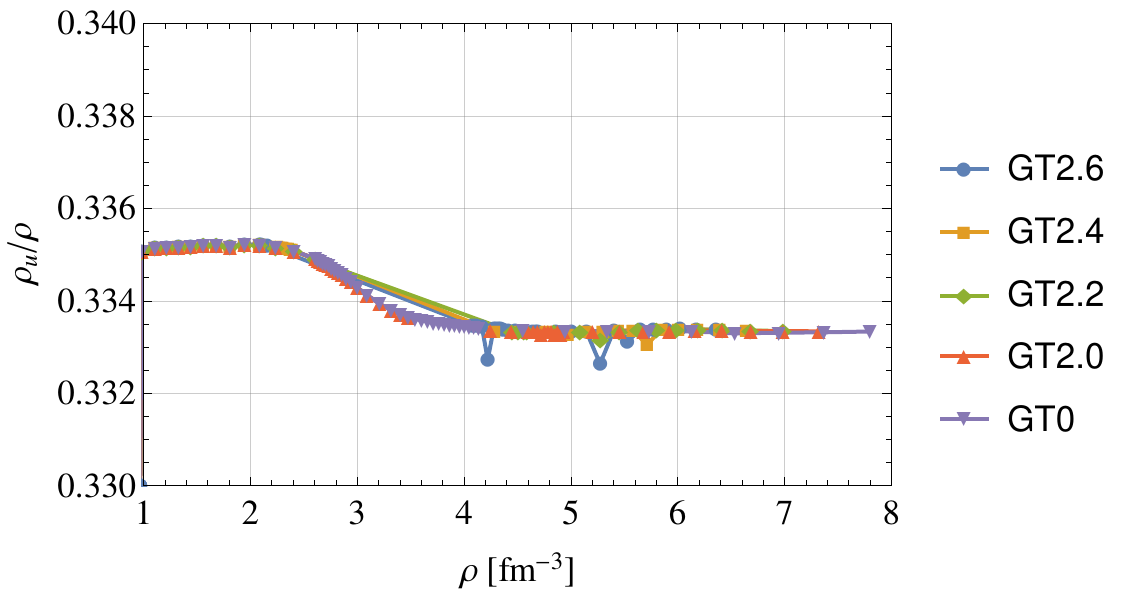}
		\caption{This figure shows the fraction of up quark number density, $\rho_u/\rho$,
as a function of total quark number density $\rho$ for each model.
			\label{fig:rhou_rho}
		}
	\end{minipage}
\\
	\begin{minipage}[t]{0.45\hsize}
		\includegraphics[height=3.5cm]{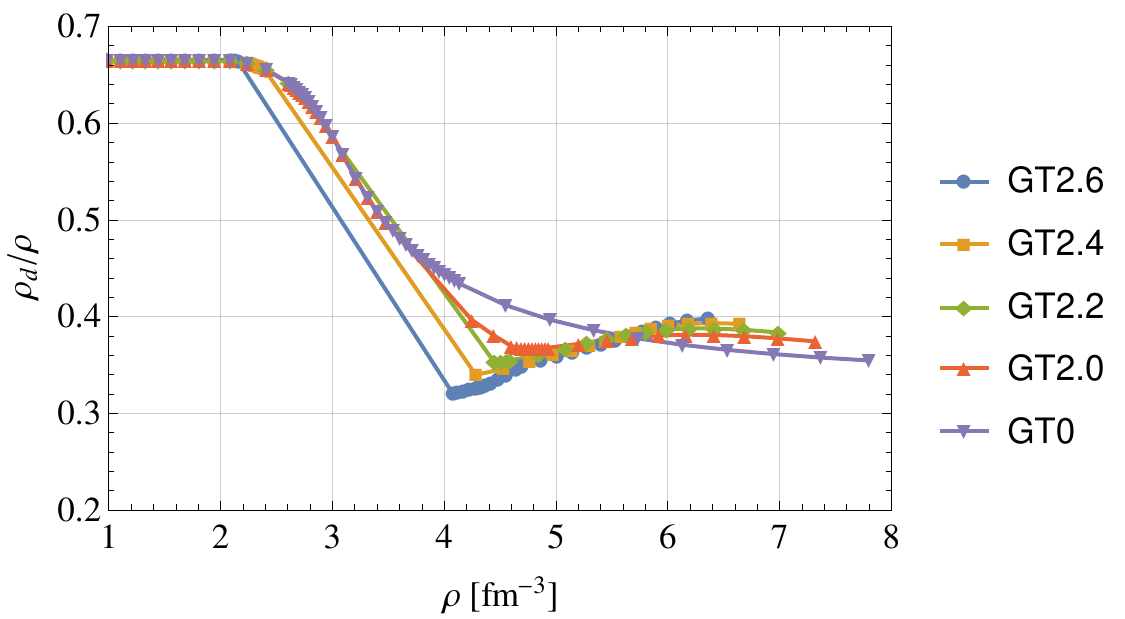}
		\caption{This figure shows the fraction of down quark number density, $\rho_d/\rho$,
as a function of total quark number density $\rho$ for each model.
			\label{fig:rhod_rho}
		}
	\end{minipage}
\qquad
	\begin{minipage}[t]{0.45\hsize}
		\includegraphics[height=3.5cm]{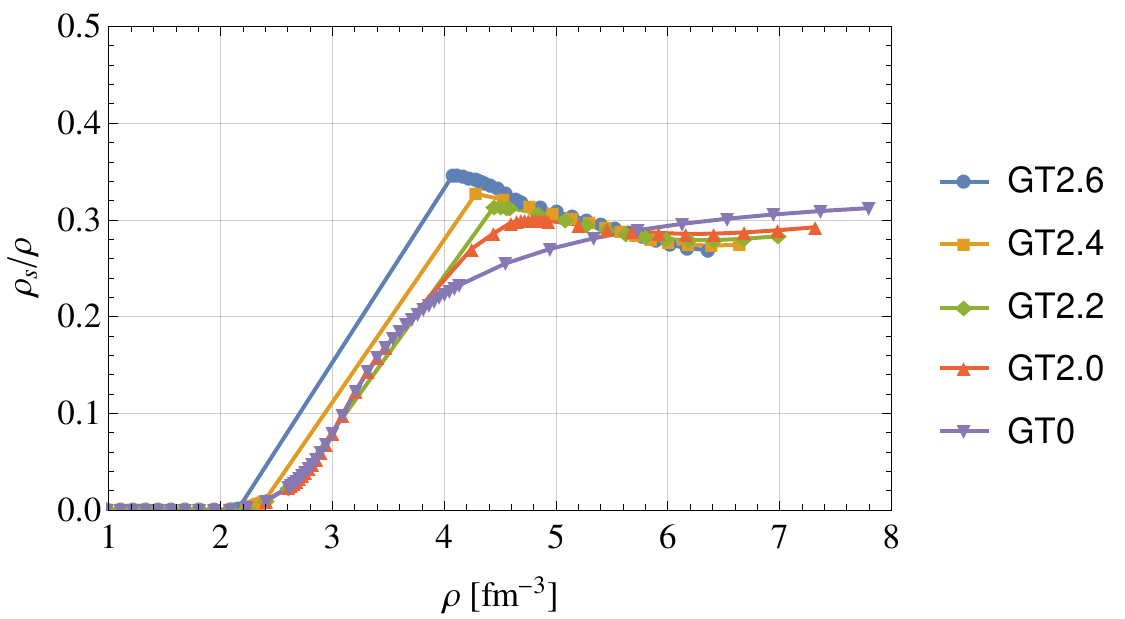}
		\caption{This figure shows the fraction of strange quark number density, $\rho_s/\rho$,
as a function of total quark number density $\rho$ for each model.
			\label{fig:rhos_rho}
		}
	\end{minipage}
\end{center}
\end{figure}

Figure \ref{fig:rhoB-mu} shows the behavior of the baryon number density $\rho_B = (\rho_u + \rho_d + \rho_s)/3$, of the system.
The horizontal and vertical axes represent the quark chemical potential and the baryon number density for each model.
The baryon number density jumps at $\mu_q \approx 0.34$ GeV for all models.
It means that the chiral restoration occurs for light quarks at this value of the quark chemical potential.
In the case of model GT0, the quark number density raises up at $\mu_q \approx 0.47$ GeV.
At this point the constituent strange quark mass starts decreasing.
As $G_T$ increases,
the quark chemical potential at which the baryon number density jumps is shifted to lower values.
These chemical potentials correspond to $\mu_{{cr_1}}$ and $\mu_{cr_{1'}}$ as is seen in Fig.\ref{fig:MqMs_mu}.
Also, the slope of the baryon number density with respect to the quark chemical potential
becomes more gentle as $G_T$ becomes larger.

Figures \ref{fig:rhou_rho}, \ref{fig:rhod_rho} and \ref{fig:rhos_rho} show the ratio of each quark
number density against the total quark number density $\rho = \rho_u + \rho_d + \rho_s$.
As is seen in Fig. \ref{fig:rhou_rho}, the proportion of up quark number density
is about $\rho_u/\rho \sim  0.334 \approx 1/3$.
Also, this figure shows that the proportion of up quark number density is almost not changed.
Compared to figures \ref{fig:rhod_rho} and \ref{fig:rhos_rho},
it is shown that the proportion of down quark number density decreases with appearance of strange quark.
As $G_T$ becomes larger,
the proportion of down/strange quark decreases/increases
in the region of $4\ \text{fm}^{-3}\leq\rho\leq 5.5\ \text{fm}^{-3}$.
On the other hand, in the larger density region,
the proportion of down/strange quark increases/decreases as $G_T$ becomes larger.
This behavior is due to an effect of the tensor interaction  which leads to
the tensor condensates.

%
\begin{figure}[t]
	\begin{center}
		\includegraphics[height=5.0cm]{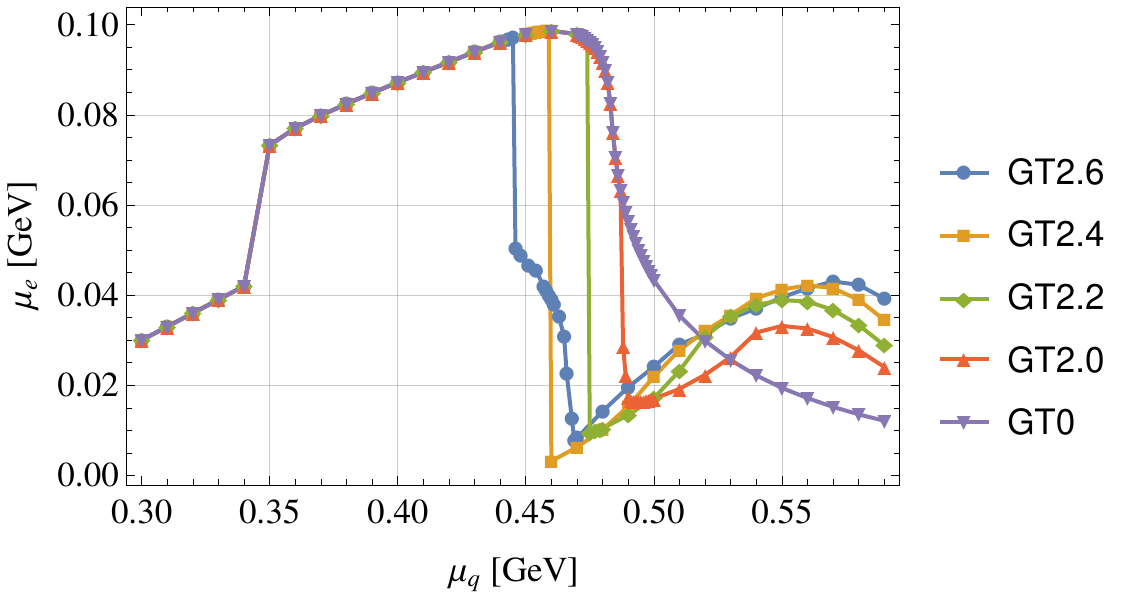}
	\end{center}
	\caption{Electron chemical potentials are depicted as a function of the quark chemical potential $\mu_q$ for each model.
		\label{fig:mue_mu}
}
\end{figure}

Figure \ref{fig:mue_mu} shows the behavior of the electron chemical potential.
The horizontal and vertical axes represent the quark chemical potential and the electron chemical potential, respectively.
In all models, $\mu_e$ has a nonzero value at small quark chemical region.
According to (\ref{eq:chargeneutrality}),
the electron chemical potential must be zero when $\rho_f = 0$.
Therefore, the small value of the electron chemical potential for $\mu_q < 0.35$ GeV may be regarded as a numerical error.
This numerical error does not affect the construction of hybrid stars in section \ref{sec:Hybridstar},
because in the low density region, we adopt the hadronic model for the equation of state.

For the Model GT0,
the electron chemical potential $\mu_e$ jumps at $\mu_q \approx 0.35$ GeV and
for 0.35 GeV $\leq \mu_q \leq 0.46$ GeV, $\mu_e$ increases
due to the appearance of the up and down quark matter.
When $\mu_q \approx 0.46$ GeV, $\mu_e$ has the maximum value
about $\mu_e \approx 0.99 $ GeV.
After overcoming the peak, $\mu_e$ decreases.
For the case of Model GT2.0/GT2.2/GT2.4,
in the region of $\mu_q \leq \mu_{cr_1}$, $\mu_e$ shows the same behavior as Model GT0.
However, at $\mu_q \approx \mu_{cr_1}$, the electron chemical potential $\mu_e$ decreases sharply.
For $\mu_{cr_1} \leq \mu_q \leq 0.55/0.55/0.56$ GeV, $\mu_e$ increases as the quark chemical potential increases.
Finally, $\mu_e$ decreases for$ \mu \geq 0.55/0.55/0.56$ GeV.
On the other hand, Model GT2.6 shows the different behavior for other models.
Similarly to other models, $\mu_e$ has the maximum value
at $\mu_q \approx 0.445 \text{GeV} (\approx \mu_{cr_{1'}})$,
and decreases at $\mu_q \approx \mu_{cr_{1'}}$.
However, this decreasing stays around $\mu_e \approx 0.05$ GeV.
In the region of $\mu_{cr_{1'}} \leq \mu_q \leq \mu_{cr_{2'}}$, the electron chemical potential $\mu_e$ decreases gradually.
At $\mu_q \approx \mu_{cr_{2'}}$, $\mu_e$ begins to increase
and it peaks at $\mu_q \approx 0.57$ GeV.

\begin{figure}[t]
	\begin{center}
		\includegraphics[height=5.0cm]{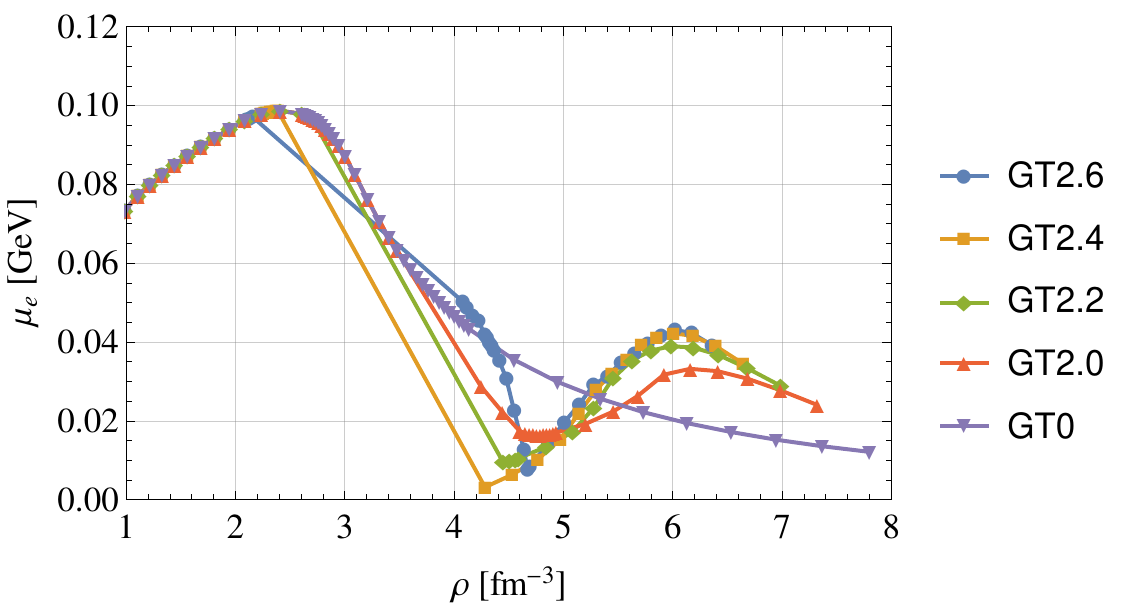}
	\end{center}
	\caption{Electron chemical potentials are depicted as a function of the total quark number density $\rho$ for each model.
		\label{fig:mue_rho}
}
\end{figure}

Figure \ref{fig:mue_rho} shows the relationship of the electron chemical potential and the total quark number density,
namely three times the baryon number density.
In this figure, it is shown that each model has a peak at $\rho \approx 2.4$ fm$^{-3}$.
The first peak arises from the appearance of the strange quark at this density.
In fact, from Fig.\ref{fig:rhos_rho}, it is seen that
the strange quark number density increases at $\rho \approx 2.4$ fm$^{-3}$.
Accordingly, the up quark number density decreases as is seen in fig \ref{fig:rhou_rho}.
Thus, as the quark or baryon number density increases, the ratio of $\rho_f / \rho$ changes,
which leads to the change of the electron chemical potential through the beta equilibrium condition.

%

\section{Hybrid star \label{sec:Hybridstar}}

The ``hybrid star" is  a kind of compact stars and
has an inner core consisting of quark matter and an outer core and crust consisting of hadrons.
In order to obtain the equation of state (EoS) of hadrons,
a density-dependent meson-nucleon couplings model with the relativistic mean-field interaction is used.
\cite{DDME2}

It is assumed that, in the inner core of the star, the quark matter exists and
the tensor condensate may appear in the high density region.
The pressure can be evaluated by the thermodynamic potential in Eq.(\ref{P}).
To obtain the inner structure of compact star,
we must solve the Toleman-Oppenheimer-Volkoff (TOV) equation from center to the outside of the star by using the EoS data:
\begin{align}
\label{eq:TOV}
	\frac{dM}{dr} &= 4\pi r^2 \rho \ ,
	\nonumber \\
	\frac{dP}{dr} &= - \frac{GM\rho}{r^2}
		\frac{
			\left(1+\frac{P}{\rho c^2}\right) \left(1+\frac{4\pi P r^3 }{M c^2}\right)
			}
		{\left(1 - \frac{2GM}{rc^2} \right)} \ ,
\end{align}
where $P$, $M$ and $\rho$ represent the pressure, mass of star and mass density profiles, respectively.
We adopt the same strategy for numerical calculation as our previous paper\cite{oursPR2}:
\begin{enumerate}
	\item{
	Give an arbitrary value to the central energy density of the hybrid star.
	}
	\item {
	If the value is large enough that the quark matter is realized, go to step 3,
	else jump to step 5.
	}
	\item{
	Solve the TOV equation from the center to the
outside of the star by using the EoS of quarks until
a certain reference pressure obtained by performing
a Maxwell construction.
	}
	\item{
	At the reference pressure,
  EoS is switched to the one of hadrons.
	}
	\item{
	Solve the TOV equation to the outside of the star
with the EoS of hadrons until the pressure of the star
vanishes.
	}
	\item{
	Change the value of central energy density and go
back to step 2.
	}
\end{enumerate}

\begin{figure}[b]
	\begin{center}
		\includegraphics[height=6.0cm]{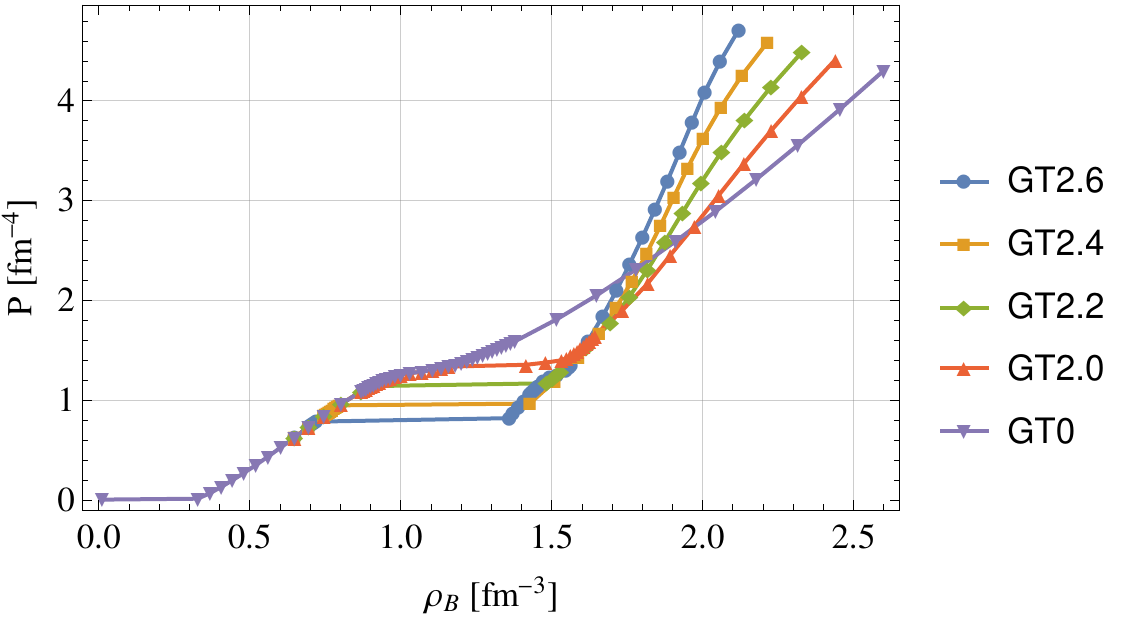}
	\end{center}
	\caption{The pressure $P$ is depicted as a function of the baryon number density $\rho_B$ for each model.
	\label{fig:P-rhoB}}
\end{figure}

In Fig.\ref{fig:P-rhoB}, the pressure is depicted as a function of the baryon number density.
In the lower density region, the crust is constructed.
As the baryon number density is increasing, the phase transition from the hadron matter to the quark matter
occurs.

\begin{figure}[b]
	\begin{center}
		\includegraphics[height=6.0cm]{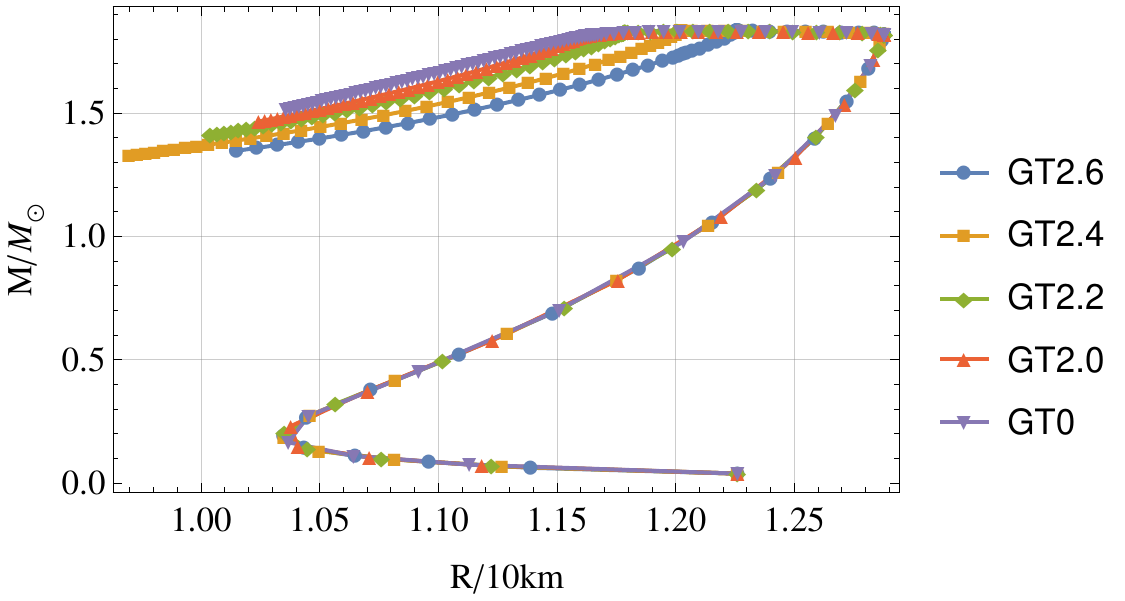}
	\end{center}
	\caption{The relationship between the radius and mass of hybrid stars is shown.
	The mass $M$ and the radius $R$ of the hybrid star are normalized by
the solar mass $M_\odot$ and the radius $R_0=10$ km, respectively.
	The horizontal and vertical axes represent the normalized radius and mass,
respectively.
	\label{fig:R_M}}
\end{figure}

Let us discuss the relationship between the radius and
mass of hybrid stars 
numerically.
Figure \ref{fig:R_M} shows the radius ($R$)-mass ($M$) relation.
In this figure, $R$ and $M$ are normalized by $R_0 = 10$ km and the solar mass $M_\odot$, respectively.
Each point corresponds to the different central energy density respectively.
First, the points are shifted from the lower right to the lower left as
the central energy density increases.
Each curve bends
at  $(R/R_0 , M/M_\odot)\approx (1.04, 0.17)$.
After that,
the points are shifted from the lower left to the upper right as the central energy density increases.
Each curve bends
at  $(R/R_0 , M/M_\odot)\approx (1.29, 1.82)$ again.
This point corresponds to
the appearance of the up and down quark matters at the inner core of
the compact star.
In the region from the lower left to the second bending point, compact stars consist of hadrons, namely, they have the hadron cores.
On the other hand, in the region from the second bending point to the upper left,
compact stars have quark cores, namely, they are hybrid stars.
In the upper left region, each curve bends again.
This point corresponds to
the appearance of the strange quarks at the inner core of
the compact star.
Specially, for the models GT2.0, GT2.2, GT2.4 and GT2.6,
this third bending point means the appearance of the tensor condensates.
As the values of $G_T$ increases,
the third bending point is shifted to right.
This shift is originated from the appearance of the strange quark at rather lower baryon number density
by the appearance of the tensor condensates.
Qualitatively, the model GT2.6 shows
some differences from the other models GT2.$\alpha$ ($\alpha \neq 6$) as for the appearance of the
tensor condensates.
However, there is no qualitative difference between the model GT2.6 and other models in the radius-mass relation.
In all models, as soon as the $F_8$ condensate sets in, the star becomes unstable.

\section{Summary and concluding remarks}

In this paper, we have investigated the behavior of the tensor
condensates and their implication to the properties of the
hybrid star by using the  Nambu-Jona-Lasinio model with the tensor-type
four-point interaction under the $\beta$ equilibrium and charge neutrality.
In the three-flavor case, it is necessary to consider the $U_A(1)$ anomaly which is incorporated
in the Kobayashi-Maskawa-'t Hooft interaction or so-called determinant interaction with six-point interaction
between quarks.

In our previous work \cite{Kagawa}, the tensor condensates have been
investigated
without charge neutrality and $\beta$ equilibrium conditions.
In that paper, the region with tensor condensates is divided into 3 parts, namely
($F_3\neq0\ ,\ F_8=0$) ,
($F_3\neq0\ ,\ F_8\neq0$) and
($F_3=0\ ,\ F_8\neq0$).
The window of the quark chemical potential where $F_3$ and $F_8$ coexist is very narrow.
On the other hand,
it is shown that, in this paper,
there is no region where only $F_3$ appears under the beta equilibrium and charge neutrality conditions.
In contrast, the region where $F_3$ and $F_8$ coexist becomes wider compared to the result of our previous paper.
This difference can be attributed to the charge neutrality and $\beta$ equilibrium conditions.
These conditions break the light quark symmetry.
If only the tensor condensate $F_3$ appears, then $F_u = |F_d|$ is realized, which leads to the light quark symmetry,
namely the up and down quarks are equivalent.
The same for the region when only $F_8$ appears.

In this paper, the hybrid stars were constructed by solving the TOV equation.
In the region of the quark matter, the equation of state was used,
which were obtained by the three-flavor NJL model imposing the
beta-equilibrium and charge neutrality conditions under the existence of the tensor condensates.
In the region of the hadron matter,
the equation of state obtained by a density-dependent meson-nucleon couplings model
with the relativistic mean-field interaction\cite{DDME2} is used.
As a result, the mass-radius relations of the hybrid stars were evaluated.
We could not obtain hybrid stars with two-solar mass
because the equation of state becomes a little too soft by the appearance of the quark matter
with the tensor condensates in the inner core of the hybrid star.
However, the repulsive interaction between quarks is not included,
and this can be the reason
that two-solar mass stars can not be hold.
It is, therefore, interesting to include the repulsive interaction in this model by introducing
the vector interaction adding to the scalar-pseudoscalar and the tensor interactions between quarks,
and
to investigate the effects of the vector interaction on the hybrid stars.
This is our next task.

In our previous work\cite{oursPR2}, the inner structure of compact stars was investigated by using of 2-flavor NJL model
with the tensor interaction.
In that paper, it was predicted that the so-called ``twin stars"\cite{twinstar1,twinstar2} may appear
by calculating the Mass-Radius relation under a rather strong coupling strength for the tensor interaction.
This occurs because the tensor condensate sets in at a smaller quark chemical potential.
In our present work, no sign of twin stars is seen.
As is seen in the result of model GT2.6,
under strong $G_T$, another tensor condensed phase appears.
Thus, in the 3-flavor NJL model we have adopted, twin stars may appear under a rather large value of $G_T$.
This will be a future investigation.

If we include the pseudovector-type four-point interaction between quarks,
the inner structure of compact star may change drastically.
By our previous work \cite{Morimoto1,Morimoto2},
the pseudovector condensates appear at a lower quark chemical potential than that appearing the tensor condensates.
Especially, the pseudovector condensates appear and vanish around the quark chemical potential where
the chiral symmetry is restored.
This region corresponds to the vicinity of the second bending point in Fig.\ref{fig:R_M}.
It is interesting to investigate the effect of pseudovector interaction on the compact stars.
This may also be one of interesting future problems.

\section*{Acknowledgements}

Two of the authors (M. M. and Y. T.) would like to express their sincere thanks
to
Professor K. Iida and Dr. E. Nakano for their helpful comments.
This work was partially supported by Fundacao para Ciencia e Tecnologia, Portugal,
under the project No. UID/FIS/04564/2020.

\end{document}